\documentclass[preprint,aps]{revtex4}
\usepackage{graphicx}
\def\Vec#1{\mbox{\boldmath $#1$}}

\newcommand{\mX}{{\Xi}}

\begin{document}

\title{
$\Lambda\Lambda$-$\Xi N$-$\Sigma\Sigma$ coupling
in $^{~6}_{\Lambda\Lambda}$He with the Nijmegen soft-core potentials
}

\author{Taiichi Yamada}
\address{Laboratory of Physics, Kanto Gakuin University,
Yokohama 236-8501, Japan
}

\date{\today}

\begin{abstract}
The $\Lambda\Lambda$-$\Xi N$-$\Sigma\Sigma$ coupling
 in $^{~6}_{\Lambda\Lambda}$He is studied with the 
 [$\alpha$+$\Lambda$+$\Lambda$] + [$\alpha$+$\Xi$+$N$] + 
 [$\alpha$+$\Sigma$+$\Sigma$] model, where the $\alpha$ 
 particle is assumed as a frozen core.
We use the Nijmegen soft-core potentials, NSC97e and NSC97f, 
 for the valence baryon-baryon part,
 and the phenomenological potentials for the $\alpha-B$ 
 parts ($B$=$N$, $\Lambda$, $\Xi$ and $\Sigma$).
We find that the calculated $\Delta B_{\Lambda\Lambda}$ 
 of $^{~6}_{\Lambda\Lambda}$He for NSC97e
 and NSC97f are, respectively, 0.6 and 0.4 MeV in the full 
 coupled-channel calculation, 
 the results of which are about half in comparison with the experimental 
 data, $\Delta B^{exp}_{\Lambda\Lambda}=1.01\pm0.20^{+0.18}_{-0.11}$ MeV. 
Characteristics of the $S=-2$ sector in the NSC97 potentials are discussed
 in detail.   

\vspace*{5mm}
\end{abstract}

\maketitle

\section{Introduction}

The study of strangeness $S=-2$ nuclei is an entrance to multistrangeness
 hadronic systems and provides unified understanding of $YN$ and $YY$
 interactions among baryon octet.
The $S=-2$ nuclei serve important information on the $YN$ and $YY$ interactions, 
 because free-space baryon-baryon scattering experiments in $S=-2$ sector 
 are difficult to be performed at the present stage.
Since the $\Lambda\Lambda$-$\Xi N$-$\Sigma\Sigma$ coupling including
 the $H$-dibaryon state is induced, there might exist exotic nuclei with $S=-2$ 
 such as $H$-nuclear states and/or hyperon-mixed nuclear states 
 among baryon octet.  

A recent discovery of ${^{~6}_{\Lambda\Lambda}{\rm He}}$ in the KEK-E373 
 experiment~\cite{Takahashi01}, which is called as {\it NAGARA}, has provided 
 a great impact in hypernuclear physics.  
The importance is due to the ambiguity-free identification of the hypernucleus and
 the high quality experimental $\Lambda \Lambda$ binding energy 
 $B_{\Lambda \Lambda}=7.25 \pm 0.19\pm ^{0.18}_{0.11}$ MeV~\cite{Takahashi01},
 which leads to a smaller $\Lambda \Lambda$ bond energy, 
 ${\Delta B_{\Lambda \Lambda}}
 ={B_{\Lambda\Lambda}({^{~6}_{\Lambda\Lambda}{\rm He}})}
 -2{B_\Lambda({^5_\Lambda{\rm He}})}=1.01 \pm 0.20 \pm ^{0.18}_{0.11}$ MeV, 
 indicating that the $\Lambda \Lambda$ interaction is more weakly attractive 
 than that reported over 30 years ago~\cite{Prowse66}.
Although it is identified as double-$\Lambda$ nucleus, we cannot exclude a possibility 
 of $H$-nuclear state or hyperon-mixed state among baryon octet, because we know 
 only the binding energy.
The data of the NAGARA event indicates four possibilities on the $YN$ and
 $YY$ interactions as follows;
1)~weakly attractive $\Lambda\Lambda$ interaction with weak 
 $\Lambda\Lambda$-$\Xi N$-$\Sigma\Sigma$ coupling effect,
2)~almost zero or weak repulsive $\Lambda\Lambda$ interaction 
 with moderate $\Lambda\Lambda$-$\Xi N$-$\Sigma\Sigma$ coupling effect,
3)~repulsive $\Lambda\Lambda$ interaction 
 with strong $\Lambda\Lambda$-$\Xi N$-$\Sigma\Sigma$ coupling effect, and
4)~$\Lambda\Lambda$-$\Xi N$-$\Sigma\Sigma$ coupling so strong as to
 produce a weakly-bound or resonant $H$-dibaryon state.
Forthcoming experiments for $S=-2$ nuclei as well as the $H$-dibaryon state
 will disclose characteristics of the $YN$ and $YY$ interactions together with 
 the structure of $S=-2$ nuclei and hyperon mixing.

After the discovery of the NAGARA event~\cite{Takahashi01}, several authors 
 have discussed the strength of the $\Lambda\Lambda$ interaction and structure of
 {$^{~6}_{\Lambda\Lambda}$He} as well as {$^{~4}_{\Lambda\Lambda}$H} 
 and {$^{~5}_{\Lambda\Lambda}$H}-{$^{~5}_{\Lambda\Lambda}$He}.
The binding energies of {$^{~6}_{\Lambda\Lambda}$He} 
 and {$^{~4}_{\Lambda\Lambda}$H} were studied with the Fadeev-Yakubovsky 
 approach~\cite{Filikhin02_1,Filikhin02_2,Filikhin03}, where they used the phenomenological 
 $\Lambda\Lambda$ interaction
 (central-type) which reproduces the low energy parameters of 
 the Nijmegen soft-core potential~\cite{Stoks99}. 
The systematic three- and four-body calculations
 of {\it p}-shell double-$\Lambda$ nuclei~\cite{Hiyama02} were performed 
 with the phenomenological $\Lambda\Lambda$ potential reproducing 
 the NAGARA data.
On the other hand, the Brueckner theory approach was applied to studying
 the $\Lambda\Lambda$ bond energy and rearrangement effect in 
 {$^{~6}_{\Lambda\Lambda}$He}.~\cite{Myint03,Kohno03,Vidana03}{\ }
The $\Xi$ component as well as the $\Lambda\Lambda$ bond energy 
 was discussed in {$^{~5}_{\Lambda\Lambda}$H} and 
 {$^{~5}_{\Lambda\Lambda}$He}~\cite{Myint03,Lanskoy04}. 
The 6-body calculation within the framework of the stochastic 
 variational method has been performed with phenomenological
 (central-type) baryon-baryon interactions.~\cite{Nemura03} 

The purpose of the present paper is to study the 
 $\Lambda\Lambda$-$\Xi N$-$\Sigma\Sigma$
 coupling effect in {$^{~6}_{\Lambda\Lambda}$He} with
 the realistic baryon-baryon potential for the two-valence-baryon part.
We use the [$\alpha$+$\Lambda$+$\Lambda$] + [$\alpha$+$\Xi$+$N$] + 
 [$\alpha$+$\Sigma$+$\Sigma$] model, where the $\alpha$ 
 particle is assumed as a frozen core. 
The Nijmegen soft-core potentials, NSC97e and NSC97f~\cite{Stoks99}, are 
 directly applied to the valence baryon-baryon interactions.
Phenomenological potentials are used for 
 the $\alpha-B$ parts ($B$=$N$, $\Lambda$, $\Xi$ and $\Sigma$).
The formulation is almost the same as one in our previous paper~\cite{Yamada00},
 where structure of light $S=-2$ nuclei and hyperon mixing were discussed.
The Pauli blocking effect of the valence nucleon in the $\alpha$+$\Xi$+$N$
 channel is taken into account properly.
We will discuss the calculated energies and coupled-channel effects
 in {$^{~6}_{\Lambda\Lambda}$He} together with the characteristics of 
 the Nijmegen potentials.  

\section{Formulation}

The total wave function of $^{~6}_{\Lambda\Lambda}$He with total angular
momentum $J$ is given as 
\begin{eqnarray}
&&\Phi=\Phi_{\Lambda\Lambda}+\Phi_{\Xi N}+\Phi_{\Sigma\Sigma},\label{total_wf}\\
&&\Phi_{\Lambda\Lambda}=\sum_{\beta=1}^{2}\sum_{LS}{\cal A}_{\Lambda\Lambda}
  \left[\Phi_{L}^{(\Lambda\Lambda)\beta}(\Vec{r}_\beta,\Vec{R}_\beta)\left[\chi_{1/2}(\Lambda)\chi_{1/2}(\Lambda)\right]_{S,I=0}\right]_J,\label{eq:LL_wf}\\
&&\Phi_{\Xi N}=\sum_{\beta=1}^{3}\sum_{LS}
  \left[\Phi_{L}^{(\Xi N)\beta}(\Vec{r}_\beta,\Vec{R}_\beta)\left[\chi_{1/2}(\Xi)\chi_{1/2}(N)\right]_{S,I=0}\right]_J,\label{eq:XN_wf}\\
&&\Phi_{\Sigma\Sigma}=\sum_{\beta=1}^{2}\sum_{LS}{\cal A}_{\Sigma\Sigma}
  \left[\Phi_{L}^{(\Sigma\Sigma)\beta}(\Vec{r}_\beta,\Vec{R}_\beta)\left[\chi_{1/2}(\Sigma)\chi_{1/2}(\Sigma)\right]_{S,I=0}\right]_J,\label{eq:SS_wf}
\end{eqnarray}
where $\beta$ denotes the Jacobian coordinate system (see Fig.~2 in Ref.~\cite{Yamada00}), 
 and $\Phi_L^{(\beta)}$ and $\chi$'s represent, respectively, the wave function of 
 the spatial part with total orbital angular momentum $L$
 and the spin-isospin functions for the valence baryons coupled to total spin $S$ and isospin $I$.
In the $\alpha$+$\Lambda$+$\Lambda$ ($\alpha$+$\Sigma$+$\Sigma$) channel,
 the antisymmetrization operator ${\cal A}_{\Lambda\Lambda}$ (${\cal A}_{\Sigma\Sigma}$)
 is needed for the two $\Lambda$ ($\Sigma$) particles.
Therefore, it is enough to take the two Jacobian coordinate systems for the
 $\alpha$+$\Lambda$+$\Lambda$ ($\alpha$+$\Sigma$+$\Sigma$) channel.
 
The wave function of the spatial part $\Phi_L(\Vec{r},\Vec{R})$ in
 Eqs.~(\ref{eq:LL_wf}), (\ref{eq:XN_wf}) and (\ref{eq:SS_wf}) is expanded 
 in terms of the Gaussian basis, which is
 known to be suited for describing both the short-range correlation and
 long-range tail behavior \cite{Kamimura88},
\begin{eqnarray}
&&\Phi_{LM}(\Vec{r},\Vec{R})=\sum_{\ell_r,\ell_R}\sum_{n_r,n_R}
C^L_{n_r\ell_r,n_R\ell_R}\left[\varphi_{\ell_r}(\Vec{r},\nu_{n_r})
\varphi_{\ell_R}(\Vec{R},\nu_{n_R})\right]_{LM},\label{eq:w.f.}\\
&&\varphi_{\ell m}(\Vec{r},\nu)=N_\ell(\nu)r^\ell\exp(-\nu r^2)Y_{\ell m}
(\hat{\Vec{r}}),\label{gs_basis}
\end{eqnarray}
where $N_\ell(\nu)$ is the normalization factor.
The Gaussian parameter $\nu$ is taken to be of geometrical progression,
\begin{equation}
\nu_n=1/b_n^2,\hspace{1cm}b_n=b_{min}a^{n-1},\hspace{1cm}n=1\sim n_{\max}.
\label{eq:para_Gaussian} 
\end{equation}
It is noted that the prescription is found to be very useful in optimizing
 the ranges with a small number of free parameters together with high
  accuracy \cite{Kamimura88}.

The total Hamiltonian within the framework of the
 [$\alpha$+$\Lambda$+$\Lambda$] + [$\alpha$+$\Xi$+$N$] + 
 [$\alpha$+$\Sigma$+$\Sigma$] model is given as 
\begin{equation}
H=\delta_{cc'}\left[T_c+V_{\alpha B_1}(\Vec{r}_1)+V_{\alpha B_2}(\Vec{r}_2)+{\Delta M}_c\right]
   +\upsilon_{cc'}(\Vec{r}_3)+\delta_{c2}V_{Pauli},\label{eq:Hamiltonian}
\end{equation}
where $c$ denotes the channel; $c=1$ for $\alpha+\Lambda+\Lambda$,
$c=2$ for $\alpha+\Xi+N$ and $c=3$ for $\alpha+\Sigma+\Sigma$.
$T_c$ and $V_{\alpha B}$ present, respectively, the kinetic energy
 operator and potential between the $\alpha$ particle and valence baryon $B$, 
 and $\upsilon_{cc'}$ denotes the interaction between the two valence baryons.
In the present study, the baryon-channel coupling is assumed to come only
 from $\upsilon_{cc'}$.
The mass difference matrix (diagonal and constant) ${\Delta M}_c$ is introduced
 to give the threshold-energy differences among the three channels, 
 ${\Delta M}_1$=0 MeV, ${\Delta M}_2$=28 MeV and ${\Delta M}_3$=160 MeV.
The Pauli principle between the $\alpha$ cluster and valence nucleon in the
 $\alpha$+$\mX$+$N$ channel is taken into account with the orthogonality
 condition model (OCM) \cite{Saito69}.
The Pauli-blocking operator $V_{Pauli}$~\cite{Kukulin84} is represented as
\begin{equation}
V_{Pauli}=\lim_{\lambda\to\infty}\lambda\mid\varphi_{0s}
(\Vec{r}_{\alpha N})\rangle\langle\varphi_{0s}(\Vec{r}_{\alpha N}')\mid,
\label{eq_A:Pauli}
\end{equation}
which removes the Pauli forbidden state $\varphi_{0s}$ between the $\alpha$ cluster 
 and valence nucleon in the core+$\mX$+$N$ three-body system.
The configuration of $\alpha$ cluster is assumed here to be of 
 simple $(0s)^4$-shell-model type.

The potential between the $\alpha$ cluster and valence hyperon $V_{\alpha Y}$
 for $Y=\Lambda$, $\Xi$ and $\Sigma$ is obtained by folding 
 the effective hyperon-nucleon ($YN$) interaction with the density of the
 $\alpha$ particle and adjusting their strength so as to reproduce 
 the experimental binding energy for the ground state of the $\alpha$+$Y$ 
 system with use of the $\alpha$+$Y$ potential model.
As for the effective hyperon-nucleon ($YN$), we use the YNG-ND 
 interaction~\cite{Yamamoto94}. 
It is known that the YNG-ND $\Lambda N$ interaction reproduces nicely 
 the $\Lambda$ binding energy of $^5_\Lambda$He as well as other light 
 $\Lambda$ hypernuclei, and the $\Xi N$ interaction is consistent with 
 the recent experimental data on $^{12}_{~\Xi}$B obtained by 
 the $^{12}$C$(K^-,K^+)$ reaction \cite{Fukuda98}.
The YNG-ND $\Sigma N$ interaction is also consistent with the experimental
 data of $^4_\Sigma$He \cite{Nagae98}. 
Concerning the density distribution for the $\alpha$ particle, we use 
 the harmonic-oscillator-type one obtained by the election
 scattering experiment \cite{Jager74}.
The $\alpha$-$\Xi$ potential obtained is so weak as to give the $\Xi$ binding
 energy as small as 0.01 MeV for the system, while
 the $\alpha$-$\Sigma$ potential produces no bound states.
As for the $\alpha$-$N$ potential, we use 
 the Kanada-Kaneko potential~\cite{Kanada79}, constructed 
 with the resonating group method (RGM) based on the microscopic theory,  
 which reproduces precisely the scattering phase shifts for
 the $p_{3/2}$, $p_{1/2}$, and $s_{1/2}$ partial waves etc.~at low energies.
The potential is local with parity dependent central and spin-orbit terms.
It is noted that we need to take into account the Pauli-blocking 
 operator in Eq.~(\ref{eq_A:Pauli}) when applying the potential to 
 the $\alpha$+$N$ ($\alpha+\Xi+N$) system.

The interaction between the two valence baryons $\upsilon_{cc'}$ in
 Eq.~(\ref{eq:Hamiltonian}) is given as
\begin{eqnarray}
\upsilon(r)&=&\upsilon^{(0)}(r)+\upsilon^{(\sigma)}(r)({\Vec{\sigma}}_1\cdot{\Vec{\sigma}}_2)
+\upsilon^{(ten)}(r)S_{12}+\upsilon^{(LS)}(r){\Vec{L}}\cdot{\Vec{S}}+\upsilon^{(ALS)}(r){\Vec{L}}
\cdot{\Vec{S}}^- \nonumber \\
&&+\upsilon^{(QLS)}(r)Q_{12}-\left[\nabla^2\phi(r)+\phi(r)\nabla^2\right],\label{BB_int}
\end{eqnarray}
where the notation is self-explanatory.
In the present paper, we use the Nijmegen soft-core potentials, 
 NSC97e and NSC97f~\cite{Stoks99}, 
 for the interaction between the two valence baryons.

The equation of motion is derived from the Rayleigh-Ritz variational method,
\begin{equation}
\delta\left[\langle {\mit \Phi} | E-H | {\mit \Phi}\rangle\right]=0.\label{eq:6_body}
\end{equation}
Solving the equation numerically, we obtain the eigenenergies of the
 Hamiltonian given in Eq.~(\ref{eq:Hamiltonian}) and expansion coefficients
 of the wave function $C$'s in Eq.~(\ref{eq:w.f.}). 

For the later discussion, it is instructive here to formulate the 
 equation of motion for the two baryon system with $S=-2$
 and spin-isospin zero.
The Hamiltonian of the system is given as
\begin{eqnarray}
h=\delta_{cc'}\left[-\frac{\hbar^2}{2\mu_c}\Vec{\nabla}^2+{\Delta M_c}\right]+\upsilon_{cc'}(\Vec{r}),
\label{H_BB}
\end{eqnarray}
where $\upsilon_{cc'}$ is the baryon-baryon interaction in Eq.~(\ref{BB_int}).
The total wave function with orbital angular momentum $\ell$ is 
presented as
\begin{eqnarray}
&&\phi=\phi_{\Lambda\Lambda}+\phi_{\Xi N}+\phi_{\Sigma\Sigma},\\
&&\phi_{\Lambda\Lambda}=\phi^{(\Lambda\Lambda)}_\ell(\Vec{r})\left[\chi_{1/2}(\Lambda)\chi_{1/2}(\Lambda)\right]_{S=I=0},\\
&&\phi_{\Xi N}=\phi^{(\Xi N)}_\ell(\Vec{r})\left[\chi_{1/2}(\Xi)\chi_{1/2}(N)\right]_{S=I=0},\\
&&\phi_{\Sigma\Sigma}=\phi^{(\Sigma\Sigma)}_\ell(\Vec{r})\left[\chi_{1/2}(\Sigma)\chi_{1/2}(\Sigma)\right]_{S=I=0},
\end{eqnarray}
where $\Vec{r}$ denotes the relative coordinate between the two-baryon system.
The wave function of the spatial part $\phi_\ell(\Vec{r})$ is expanded into the Gaussian basis
 $\varphi_\ell$ in Eq.~(\ref{gs_basis}),
\begin{eqnarray}
\phi_\ell(\Vec{r})=\sum_n c_{n\ell}\varphi_\ell(\Vec{r},\nu_n).\label{wf_BB}
\end{eqnarray}
The equation motion is derived from the variational method, 
 $\delta\left[\langle\phi|\epsilon-h|\phi\rangle\right]=0$, 
 and corresponds to that in Eq.~(\ref{eq:6_body}) under 
 the condition of choosing only the Jacobian coordinate with 
 $\beta=1$ in Eqs.~(\ref{eq:LL_wf})$\sim$(\ref{eq:SS_wf}) and setting to 
 $|\Vec{R}_1|\rightarrow\infty$, where $\Vec{R}_1$ denotes the relative coordinate
 of the $\alpha-(BB)$ part with $(BB)=(\Lambda\Lambda)$, ($\Xi N$) and ($\Sigma\Sigma$).

\section{Results and discussion}

\subsection{${^1S_0}$ state of two-baryon system with ${S=-2}$ and ${I=0}$}

It is instructive, first of all, to study the characteristics of the NSC97e 
 and NSC97f potentials by solving the Schr\"odinger equation for the two-baryon 
 system with $S=-2$ and isospin $I=0$ with the Hamiltonian in Eq.~(\ref{H_BB}).
The calculated energies of the $^1S_0$ state, 
 $E_{\Lambda\Lambda}$ $(=-B_{\Lambda\Lambda})$, with respect to the
 $\Lambda+\Lambda$ threshold are listed in Table~\ref{tab:1} 
 for various coupled-channel cases.

In the single $\Lambda\Lambda$ channel case, 
 both the NSC97e and NSC97f potentials give no bound states as well as in 
 the $\Lambda\Lambda$ and $\Xi N$ coupled-channel case.
This result suggests that the coupling effect of the $\Lambda\Lambda$-$\Xi N$ 
 conversion potential is weak in both the potentials.
Figure~\ref{fig:1} shows the radial behaviors of 
 $\upsilon_{\Lambda\Lambda-\Lambda\Lambda}(r)$, 
 $\upsilon_{{\Xi N}-{\Xi N}}(r)$ and $\upsilon_{\Lambda\Lambda-{\Xi N}}(r)$ 
 for NSC97e, where the momentum-dependent term
 in Eq.~(\ref{BB_int}) is not included.
The behaviors for NSC97f are almost the same as those for NSC97e.
Although the coupling potential, $\upsilon_{\Lambda\Lambda-{\Xi N}}$, 
 is strong in the short-range region, the coupling effect becomes weak
 due to the following reasons:~  
Both the $\upsilon_{\Lambda\Lambda-\Lambda\Lambda}$ 
 and $\upsilon_{{\Xi N}-{\Xi N}}$ potentials have very strongly repulsive part 
 in the short-range region and very weakly attractive one in the outer 
 region, the characteristics of which make the amplitude of the relative
 wave function between the two $\Lambda$ particles (as well as between the
 $\Xi$ and $N$ particles) smaller,
 so that the coupling matrix element becomes very small.
Consequently, the coupling effect is weak in the $\Lambda\Lambda$-$\Xi N$
 channel system.
 
In case of the $\Lambda\Lambda$ and $\Sigma\Sigma$ coupled-channel problem
 with NSC97e, however, we find that a bound state 
 with $B_{\Lambda\Lambda}\sim 21$ MeV~\cite{Rijken03},
 where the channel components are $P_{\Lambda\Lambda}=48 \%$ and
 $P_{\Sigma\Sigma}=52 \%$ (see Table~\ref{tab:1}).
The result is surprising for us.
The mechanism of producing such a bound state is follows:~
Figure~\ref{fig:2} shows the radial behaviors of 
 $\upsilon_{\Lambda\Lambda-\Lambda\Lambda}(r)$, 
 $\upsilon_{\Sigma\Sigma-\Sigma\Sigma}(r)$ and 
 $\upsilon_{\Lambda\Lambda-\Sigma\Sigma}(r)$ for NSC97e,
 where the momentum-dependent term in Eq.~(\ref{BB_int}) is not included.
The attractive (repulsive) behavior of $\upsilon_{\Sigma\Sigma-\Sigma\Sigma}$ 
 ($\upsilon_{\Lambda\Lambda-\Lambda\Lambda}$) in the short-range region
 makes the energy of the $\Sigma\Sigma$-channel ($\Lambda\Lambda$-channel) 
 state push down (push up).
Both the energies, then, are almost degenerate
 in energy or the energy of the $\Sigma\Sigma$ channel is slightly smaller than
 that of the $\Lambda\Lambda$ channel.
Reflecting the very strong coupling potential $\upsilon_{\Lambda\Lambda-\Sigma\Sigma}$
 as shown in Fig.~\ref{fig:2}, then, the bound state appears in the short-range region.
Although the above-mentioned explanation is a little bit schematic,
 the bound state is produced dynamically in the short-range region
 by solving the $\Lambda\Lambda$ and $\Sigma\Sigma$ coupled-channel problem
 in which the momentum-dependent term is switched on. 
On the other hand, in case of NSC97f, we could not find such a bound state 
 in NSC97e (see Table~\ref{tab:1}).
The reasons are follows:~
Although the radial behavior of the coupling potential 
 $\upsilon_{\Lambda\Lambda-\Sigma\Sigma}$ in NSC97f 
 is almost the same as that in NSC97e, we found significantly quantitative differences
 of the diagonal potentials, $\upsilon_{\Sigma\Sigma-\Sigma\Sigma}$ and
 $\upsilon_{\Lambda\Lambda-\Lambda\Lambda}$, between NSC97e and 97f:~
The strength of the attraction (the repulsion) of $\upsilon_{\Sigma\Sigma-\Sigma\Sigma}$ 
 ($\upsilon_{\Lambda\Lambda-\Lambda\Lambda}$) in NSC97f 
 decreases (increases) by about $50\sim70$ \% (by about 25 \%) 
 in the short-range region in comparison with NSC97e. 
The NSC97f potential, thus, has no ability to produce the situation like NSC97e,  
 that the $\Sigma\Sigma$- and $\Lambda\Lambda$-channel 
 states are almost degenerate in energy as discussed above,
 so that there is no bound state in the $\Lambda\Lambda$-$\Sigma\Sigma$ 
 channel system in NSC97f.  
The strong coupling effect with the $\Sigma\Sigma$ channel in NSC97f, however, can
 be seen in the $\Xi N$-$\Sigma\Sigma$ channel which has a bound state
 with the binding energy $B_{\Lambda\Lambda}(=B_{\Xi N}-28)\sim$82 MeV (see Table~\ref{tab:1}).
Such a bound state appears also in NSC97e, but its binding energy is $B_{\Xi N}\sim$24 MeV.
We observe considerable quantitative differences of the binding 
 energies of the two-baryon systems between NSC97e and 97f. 
Although the existence of the bound states in the two-channel problems 
 seems to be strange, it is related to a deeply bound state in the
 $\Lambda\Lambda$-$\Xi N$-$\Sigma\Sigma$ channel, the details of which
 will be discussed below. 

Switching on the $\Lambda\Lambda$-$\Xi N$-$\Sigma\Sigma$ coupling,
 we find an extremely deeply bound state with the binding energy 
 of $B_{\Lambda\Lambda}$(=$-E_{\Lambda\Lambda}$)=1475 (1624) MeV 
 for NSC97e (NSC97f), where the channel components are 
 $P_{\Lambda\Lambda}$=26.6 (23.9) \%, $P_{\Xi {\rm N}}$=31.1 (32.1) \% 
 and $P_{\Sigma\Sigma}$=42.4 (44.0) \%~\cite{Rijken03}.
The appearance of the bound states astonishes us.
It is noted that the momentum-dependent term in Eq.~(\ref{BB_int})
 is taken into account in the calculation.
Although the calculated binding energies depend slightly on the choice
 of the Gaussian size parameters in Eq.~(\ref{wf_BB}) because of a
 singularity of the Nijmegen potentials at origin ($r=0$)~\cite{Rijken03},
 the qualitative characteristics of the deeply bound state do not change very much.
The radial wave function of the deeply bound state in the full coupled-channel case 
 for NSC97e is illustrated in Fig.~\ref{fig:3}.
The behavior of the wave function shows that the bound state is localized
 at $r\leq 0.5$ fm.  
From the channel components for the state (see Table~\ref{tab:1}) and relative 
 phases among the three channels (see Fig.~\ref{fig:1}),  the deeply bound state obtained
 has a flavor-SU(3)-$\{8s\}$-like character, $|{8s}\rangle=\sqrt{1/5}|\Lambda\Lambda\rangle
 +\sqrt{1/5}|\Xi{\rm N}\rangle+\sqrt{3/5}|\Sigma\Sigma\rangle$.
It is interesting, here, to study the effect of the momentum-dependent (MD) term
 on the binding energy and channel components of the deeply bound state.
The calculated binding energy without MD is as large as 
 $B_{\Lambda\Lambda}$=2088 (2374) MeV for NSC97e (97f), 
 and thus, we found that the repulsive effect of MD is as large as 
 about $600\sim700$ MeV, whose repulsive character can be inferred 
 from the definition [see Eq.~(\ref{BB_int})].
On the other hand, the resultant channel components without the MD term
 are almost  the same as those with it.
The results, thus, indicate that the deeply bound state is not produced
 by only the MD term.

In order to study the mechanism of appearing the deeply bound state, 
 the effective potential defined in Ref.~\cite{Maessen89} is 
 calculated for the flavor-SU(3)-$\{8s\}$ state,  
 the result of which is illustrated in Fig.~\ref{fig:4} 
 together with the flavor-SU(3)-$\{27\}$ potential for reference.
In the calculation, the momentum-dependent (MD) term in Eq.~(\ref{BB_int}) is 
 explicitly taken into account.
Although the flavor-SU(3)-$\{27\}$ potential behaves normally, we find a strong attraction 
 in the short-range region for the flavor-SU(3)-$\{8s\}$ potential which produces
 the flavor-SU(3)-$\{8s\}$-like deeply bound state in the NSC97e potential.
The results for NSC97f are almost the same as those for NSC97e. 
In case without the MD term, the radial behavior of the flavor-SU(3)-$\{27\}$ 
 potential is qualitatively similar to that with the MD term, but the strength of 
 the attraction in the short-range region for the former is much larger than 
 that for the latter. 
Thus, the origin of the deeply bound state comes mainly from the fact that 
 the bare terms without the MD term in Eq.~(\ref{BB_int}) are designed to have 
 much strong attraction in the flavor-SU(3)-$\{8s\}$ channel, the strength of 
 which is so large as to overcome the repulsive effect originating from the MD term.

According to the quark-cluster model, one Pauli-forbidden state, 
 a flavor-SU(3)-$\{8s\}$ state with the quark shell-model ${(0s)}^6$ configuration, 
 appears in the $^1S_0$ two-baryon system with $S=-2$ and $I=0$ \cite{Oka84}.
The deeply bound states in NSC97e and NSC97f, thus, might be regarded 
 as playing a Pauli-forbidden-state-like role in the potentials, although
 we don't know any reasons of why such a bound state is incorporated
 in the Nijmegen OBEP framework.
It is remarked that the existence of the deeply bound states does not affect 
 to the calculated results of the low-energy scattering parameters
 of NSC97e and NSC97f~\cite{Stoks99}, because of the binding
 energy as large as $1500\sim1600$ MeV.

In this paper, we call the deeply bound states observed here as
 {\it pseudo bound states}, and impose the following condition 
 on the wave function: physical states should be orthogonal 
 to the pseudo bound states, 
 when the NSC97 potentials are used in theoretical calculations.
Since the $0^+_2$ state obtained in the $\Lambda\Lambda$-$\Xi N$-$\Sigma\Sigma$ 
 coupled-channel calculation in Table~\ref{tab:1} is orthogonal to the pseudo bound state
 ($0^+_1$), the $0^+_2$ state is interpreted as a physical state.
The calculated results with the single channel problems 
 and the two-channel problems in Table~\ref{tab:1}, thus, lose physical meanings 
 because their wave functions are not orthogonal to the {\it pseudo bound states}. 
We should note that the orthogonal condition does not give any influence 
 to the results of the scattering length and effective range 
 of NSC97e and NSC97f~\cite{Stoks99}.

Careful treatments are needed to perform the structure calculation of 
 $^{~6}_{\Lambda\Lambda}$He, where the total wave function should not contain 
 any component of the pseudo bound state.
In the present framework, we can easily remove the pseudo-bound-state 
 component from the total wave function in Eq.~(\ref{total_wf}) by introducing the following 
 exclusion operator in the total Hamiltonian in Eq.~(\ref{eq:Hamiltonian}),
\begin{eqnarray}
V_{PBS}=\lim_{{|\lambda'|}\to\infty}{\lambda'}\mid\varphi_{PBS}
(\Vec{r}_{BB})\rangle\langle\varphi_{PBS}(\Vec{r}_{BB}')\mid,\label{PBS}
\end{eqnarray}  
where $\varphi_{PBS}(\Vec{r}_{BB})$ is the pseudo bound state 
 of the two-baryon system with the flavor-SU(3)-$\{8s\}$-like
 character as mentioned above and $\Vec{r}_{BB}$ denotes the relative 
 coordinate between the two baryons.
The results of $^{~6}_{\Lambda\Lambda}$He will be presented 
 in the next subsection (Sec.~IIIB).

\subsection{$\Lambda\Lambda$-$\Xi N$-$\Sigma\Sigma$ coupling in {$^{~6}_{\Lambda\Lambda}$He}} 

Table~\ref{tab:2} shows the full $\Lambda\Lambda$-$\Xi N$-$\Sigma\Sigma$ 
 coupled-channel results of {$^{~6}_{\Lambda\Lambda}$He} with the
 exclusion operator $V_{PBS}$ in Eq.~(\ref{PBS}) which removes the pseudo-bound-state 
 component from the total wave function of {$^{~6}_{\Lambda\Lambda}$He}.
The calculated $\Delta B_{\Lambda\Lambda}$ of {$^{~6}_{\Lambda\Lambda}$He} 
 is 0.61 (0.36) MeV for NSC97e (NSC97f), the value of which is about half 
 in comparison with the experimental data, 
 $\Delta B^{exp}_{\Lambda\Lambda}=1.01\pm0.20^{+0.18}_{-0.11}$ MeV.  
The $\Lambda\Lambda$, $\Xi N$ and $\Sigma\Sigma$ components are, respectively,
 $P_{\Lambda\Lambda}$=99.77 (99.81)~\%, $P_{\Xi N}$=0.21 (0.18)~\% and
 $P_{\Sigma\Sigma}$=0.01 (0.01)~\% for NSC97e (NSC97f).
The results indicate that the hyperon mixing effect is very small in the ground state
 of {$^{~6}_{\Lambda\Lambda}$He}.
Although the Nijmegen potentials have a singularity at the origin as mentioned 
 in Sec.~IIIA, we found that the effect to the calculated binding 
 energy is less than 0.02 MeV because we used the exclusion operator $V_{PBS}$.

Here, it is instructive to discuss the calculated results of {$^{~6}_{\Lambda\Lambda}$He} 
 for various coupled-channel cases without the exclusion operator $V_{PBS}$,
 although we have to remind that only the results of the full coupled-channel
 calculations with $V_{PBS}$ are {\it physical} in the present paper
 as discussed in Sec.~IIIA.
The results are shown in Table~\ref{tab:2}.

First let us see the results of only the $\Lambda\Lambda$ channel
 switching off the couplings with the $\Xi N$ and $\Sigma\Sigma$ channels.
The calculated $\Delta B_{\Lambda\Lambda}$ of
 {$^{~6}_{\Lambda\Lambda}$He} is as small as 0.33 and 0.09 MeV,
 respectively, for NSC97e and NSC97f.
In case of the $\alpha$+$\Lambda$+$\Lambda$ and 
 $\alpha$+$\Xi$+$N$ coupled-channel problem, we find that 
 $\Delta B^{cal}_{\Lambda\Lambda}$ is 1.36 (0.56) MeV
 for NSC97e (NSC97f), where the respective $\Xi N$-channel
 components are as small as 0.55 (0.21) \%.
Switching on the full $\Lambda\Lambda$-$\Xi N$-$\Sigma\Sigma$ coupling,
 we obtain a deeply bound state of {$^{~6}_{\Lambda\Lambda}$He}, $0^+_1$,  
 with the flavor-SU(3)-$\{8s\}$-like character
 for both the NSC97e and NSC97f potentials (see Table~\ref{tab:2}).
This is due to the fact that the two-baryon system ($\Lambda\Lambda$-$\Xi N$-$\Sigma\Sigma$)
 with $S=-2$ and $I=0$ has a deeply bound state with the flavor-SU(3)-$\{8s\}$-like character
 as discussed in Sec.~IIIA.
The reason of why the $\Delta B_{\Lambda\Lambda}$ values of $0^+_1$ 
 in Table~\ref{tab:2} are smaller than those in the two-baryon system in Table~\ref{tab:1}
 is ascribed mainly to the Pauli-blocking effect for the $\alpha+\Xi+N$ channel.
The deeply bound state ($0^+_1$) of {$^{~6}_{\Lambda\Lambda}$He}, thus,  
 corresponds to a pseudo bound state or an unphysical state in the present study, 
 and the second $0^+$ state corresponds to a candidate of the physical state, 
 although the state may have some component of the pseudo bound state of
 the two-valence-baryon system.
The calculated $\Delta B_{\Lambda\Lambda}$ of the $0^+_2$ state 
 is 0.61 (0.36) MeV for NSC97e (NSC97f) in Table~\ref{tab:2}.
The values as well as the channel components are exactly the same as those 
 with the exclusion operator $V_{PBS}$.
The reason is due to the fact that the binding energy of the pseudo bound state
 of the two-baryon system with $S=-2$ and $I=0$ 
 is as large as $B_{\Lambda\Lambda}=1500\sim1600$ MeV, and
 therefore, the existence of the pseudo bound state gives 
 almost no effect to the $0^+_2$ state of {$^{~6}_{\Lambda\Lambda}$He}.
Comparing the energies of the $0^+_2$ state in the full coupled-channel 
 problem with those of the $\alpha$+$\Lambda$+$\Lambda$ and 
 $\alpha$+$\Xi$+$N$ coupled-channel one,
 we observe that the $\Delta B^{cal}_{\Lambda\Lambda}$ value becomes
 smaller due to the $\Sigma\Sigma$-channel coupling.
Although it seems that the reduction suggests a repulsive effect of 
 the $\Sigma\Sigma$ channel, the origin of the reduction is due to 
 the existence of the pseudo bound state.

From the above results together with those in Sec.~IIIA, we learn that 
 the theoretical discussion on the binding energy of $S=-2$ nuclei
 without the $\Lambda\Lambda$-$\Xi N$-$\Sigma\Sigma$
 coupling has no definite sense, because of the existence of the pseudo bound state 
 in the NSC97e and NSC97f potentials.
The pseudo bound state comes only from the full-channel calculation and is
 not produced in the single or two-channel problems.
In the present framework, the component of the pseudo bound state is removed from
 the total wave function of {$^{~6}_{\Lambda\Lambda}$He} with the exclusion 
 operator in the full coupled-channel calculation.
The existence of the pseudo bound state, thus, enforces us to discuss only 
 the full-channel problem, and it is not adequate to compare directly 
 the results of the single-channel and two-channel calculations 
 with those of the full-channel one. 

It is interesting to see the effects of the binding energy and channel
 components of {$^{~6}_{\Lambda\Lambda}$He} on the choice of $\alpha$-$B$ 
 potentials ($B$ denotes baryon).
The calculated results are discussed, hereafter, in the three-channel coupled 
 problem including $V_{PBS}$ [Eq.~(\ref{PBS})] with NSC97e 
 for the following four cases;~1)~using the folding-type $\alpha$-$N$ 
 potential~\cite{Yamada00} , 
 2)~neglecting the Pauli-blocking operator in Eq.~(\ref{H_BB}), 
 3)~neglecting the $\alpha$-$\Xi$ potential in the present study 
 and 4)~using the strongly repulsive 
 $\alpha$-$\Sigma$ potential suggested in the experimental analysis of
 the $^{28}$Si($\pi^-,K^+$) reaction~\cite{Noumi02}.

First we study the effect on the choice of the $\alpha$-$N$ potential.
Although the Kanada-Kaneko (KK) potential was used in the present study,
 it is instructive to apply the folding potential which had used in the previous 
 our paper~\cite{Yamada00}.
It is derived from the folding procedure of the effective $NN$ interaction, 
 HNY~\cite{Hasegawa71}, with the density of the $\alpha$ particle.
The calculated results are follows:~$\Delta B_{\Lambda\Lambda}$=0.60 MeV,
 $P_{\Lambda\Lambda}$=99.79~\%, $P_{\Xi N}$=0.19~\% and
 $P_{\Sigma\Sigma}$=0.01~\%.
They are almost the same as those with the KK potential (see Table~\ref{tab:2}).
We found, thus, that the calculated results do not depend on the details
 of the $\alpha$-$N$ potentials very much, because of the
 $\Xi N$ component as small as 0.2~\% in {$^{~6}_{\Lambda\Lambda}$He.
Secondly, the effect of imposing the Pauli-blocking operator
 in Eq.~(\ref{eq_A:Pauli}) for the $\alpha$-$N$ system is investigated by dropping it 
 out of the structure calculation for {$^{~6}_{\Lambda\Lambda}$He}. 
In that case, both the $\Xi$ particle and valence nucleon ($N$) can be in $S$ orbit 
 and the nucleon is bound by 12 MeV with respect to the $\alpha$+$n$ threshold 
 for the KK potential. 
We expect that the $\Xi N$ component is enhanced.
The results are follows: $\Delta B_{\Lambda\Lambda}$=0.81 MeV,
 $P_{\Lambda\Lambda}=99.40$~\%,  $P_{\Xi N}=0.58$~\% and  
 $P_{\Sigma\Sigma}=0.02$~\%.
We see that the component of the $\Xi N$ channel without
 the Pauli-blocking operator is about three times 
 larger than that with the operator (see Table~\ref{tab:2}), 
 although the energy gain is as small as about 0.2 MeV.
This result encourages us to expect that the mass $A=5$ system with $S=-2$
 has the $\Xi N$ component larger than the present $A=6$ system,
 because the former has no Pauli-blocking effect in the $\Xi$ 
 channel~\cite{Myint03,Lanskoy04}.

Thirdly, the effect on the $\alpha$-$\Xi$ potential is investigated.
In the present study, we used the folding potential derived from folding
 the YNG-ND $\Xi N$ interaction with the density of the $\alpha$ particle,
 which gives the $\Xi$-particle binding energy as small as $B_{\Xi}$=0.01 MeV
 for the $\alpha$+$\Xi$ system.
Although the $\Xi N$ interaction is consistent with the recent experimental 
 data on $^{12}_\Xi$B produced in the $^{12}$C($K^-,K^+$) reaction,
 it is interesting to study what happens if the weak $\alpha$-$\Xi$ potential
 is neglected.  
The results are follows: $\Delta B_{\Lambda\Lambda}$=0.60 MeV,
 $P_{\Lambda\Lambda}=99.79$~\%,  $P_{\Xi N}=0.20$~\% and  
 $P_{\Sigma\Sigma}=0.01$~\%.
They are almost the same as those with switching on the $\alpha$-$\Xi$ potential
 (see Table~\ref{tab:2}).
Thus, the effect from the $\alpha$-$\Xi$ potential is very weak 
 in the ground state of {$^{~6}_{\Lambda\Lambda}$He}, 
 reflecting the very small component of the $\Xi N$ channel.

Finally, we study the effect on the $\alpha$-$\Sigma$ potential.
The folding-type $\alpha$-$\Sigma$ potential was applied in Table~\ref{tab:1},
 where we used the YNG-ND $\Sigma N$ interaction
 which is consistent with the experimental data of $^4_\Sigma$He. 
The $\alpha$-$\Sigma$ potential has a weak repulsive character. 
The recent experimental data on the $^{28}$Si$(\pi^-,K^+)$ reaction~\cite{Noumi02}, 
 however, suggested that a strongly repulsive $\Sigma$-nucleus potential is needed 
 to reproduce the observed spectrum within the framework of DWIA.
The real part of the phenomenological potential is of the Woods-Saxon (WS) type,
 $U(r)=U_0/\{1+\exp[(r-c)/z]\}$, with $U_0$=150 MeV,
 $c=1.1\times(A-1)^{1/3}$ fm and $z=0.67$ fm~\cite{Noumi02}. 
It is interesting to investigate what happens if we use 
 such a strong repulsive WS-type potential in {$^{~6}_{\Lambda\Lambda}$He}.
The results with $V_{PBS}$ in Eq.~(\ref{PBS}) are 
 follows:~$\Delta B_{\Lambda\Lambda}$=0.59 MeV,
 $P_{\Lambda\Lambda}=99.78$~\%,  $P_{\Xi N}=0.21$~\% and  
 $P_{\Sigma\Sigma}=0.01$~\%.
They are almost the same as those in Table~\ref{tab:2}.
The reason is due to the extremely small $\Sigma\Sigma$ component
 ($P_{\Sigma\Sigma}=0.01$~\%) in the ground state 
 of {$^{~6}_{\Lambda\Lambda}$He} (see Table~\ref{tab:2}).
Some effects, however, were found in the pseudo-bound state (PBS) of 
 {$^{~6}_{\Lambda\Lambda}$He} in the calculation 
 without $V_{PBS}$:~$\Delta B_{\Lambda\Lambda}$=1460 MeV,
 $P_{\Lambda\Lambda}=26.54$~\%,  $P_{\Xi N}=30.86$~\% and  
 $P_{\Sigma\Sigma}=42.60$~\%.
Comparing with the results in Table~\ref{tab:2}, the binding energy
 decreases by about 7 MeV reflecting the repulsive character
 of the WS-type potential.
The results indicate that the WS-type $\alpha$-$\Sigma$
 potential is not so strongly repulsive as to suppress the PBS.
The reason of why the potential does not suppress the PBS is given 
 as follows: The PBS in {$^{~6}_{\Lambda\Lambda}$He} has a main structure 
 of $\alpha$+$"BB"$, where $"BB"$ denotes the PBS in the two-baryon system 
 with $S=-2$ with $B_{\Lambda\Lambda}$=1475 MeV. 
The main part of the binding energy 
 in {$^{~6}_{\Lambda\Lambda}$He}(PBS), thus, comes from
 the interaction energy between the two valence baryons.
This means that the contribution from the $\alpha$-$B$ potentials
 is not so large in comparison with that from the intra valence-baryon
 interactions.
In addition, the effect of the WS-type repulsive $\alpha$-$\Sigma$ 
 potential is weakened in {$^{~6}_{\Lambda\Lambda}$He}(PBS)
 because the $\Sigma\Sigma$ component is only about 40 \%. 
Consequently, the binding energy of PBS 
 in {$^{~6}_{\Lambda\Lambda}$He} is not changed drastically even though we use 
 the repulsive WS-type $\alpha$-$\Sigma$ potential.

\section{Summary}

The $\Lambda\Lambda$-$\Xi N$-$\Sigma\Sigma$ coupling
 in $^{~6}_{\Lambda\Lambda}$He was investigated with the 
 [$\alpha$+$\Lambda$+$\Lambda$] + [$\alpha$+$\Xi$+$N$] + 
 [$\alpha$+$\Sigma$+$\Sigma$] model, where the $\alpha$ 
 particle is assumed as a frozen core.
The Nijmegen soft-core potentials, NSC97e and NSC97f, 
 were used for the valence baryon-baryon part,
 and the phenomenological potentials were employed 
 for the $\alpha-B$ parts ($B$=$N$, $\Lambda$, $\Xi$ and $\Sigma$).
In the two-baryon system ($\Lambda\Lambda$-$\Xi N$-$\Sigma\Sigma$)
 with ${^1S_0}$ and $I=0$, we found that the NSC97e and NSC97f 
 have a deeply bound state ($B_{\Lambda\Lambda}=1500\sim1600$ MeV), 
 whose character is of the flavor-SU(3)-\{8s\}-like
 and is similar to the Pauli-forbidden state in the quark-cluster model.
Such a deeply bound state gives no effect
 to the low-energy scattering parameters of NSC97e (NSC97f), 
 although the existence seems to be improper in the OBEP framework. 
We thus called it as a pseudo bound state and 
 impose the following condition on the wave function 
 of many baryon system when the NSC97 potentials are applied to 
 many-body system: physical states should be 
 orthogonal to the pseudo bound states. 
It was then found that the calculated $\Delta B_{\Lambda\Lambda}$ of
 $^{~6}_{\Lambda\Lambda}$He for NSC97e and NSC97f are, 
 respectively, 0.6 and 0.4 MeV in the full coupled-channel calculation.
The results are about half in comparison with the experimental data, 
 $\Delta B^{exp}_{\Lambda\Lambda}=1.01\pm0.20^{+0.18}_{-0.11}$ MeV. 

In the present study, we neglected the $\Sigma$-$\Lambda$ coupling 
 effect, the importance of whose effect has been recently pointed out 
 in $^5_\Lambda$He~\cite{Nemura02}.
Since the $^5_\Lambda$He nucleus is expected to be described with
 the extended $\alpha$-cluster model, $[(3N+N)+\Lambda]$ and 
 $[(3N+N)+\Sigma]$, we can study the $\Lambda$-$\Sigma$ coupling effect 
 in $^{~6}_{\Lambda\Lambda}$He, if we use the $[(3N+N)+\Lambda+\Lambda]$
 + $[(3N+N)+\Xi+N]$ + $[(3N+N)+\Sigma+\Sigma]$ model.
The study will give a basic starting point for us to perform the systematic 
 structure study taking into account the $\Lambda$-$\Sigma$ coupling 
 in $p$-shell double-$\Lambda$ hypernuclei as well as the single-$\Lambda$ 
 hypernuclei. 
Such an approach is now planning and partially in progress.

\section*{Acknowledgments}
We greatly acknowledge helpful discussions with E.~Hiyama, M.~Kamimura, T.~Motoba,
 Th.A.~Rijken, and Y.~Yamamoto.

\clearpage
\begin{table}
\caption{
Calculated energies ($E_{\Lambda\Lambda}=-B_{\Lambda\Lambda}$) 
and channel components of the $^1S_0$ state 
of two-baryon system with $S=-2$ and $I=0$
for various coupled channel cases, where we use the NSC97e and
NSC97f potentials with the momentum-dependent term.
}
\label{tab:1}
\begin{center}
\begin{tabular}{cccccc}
\hline
\hline
\multicolumn{6}{c}{NSC97e potential}\\
\hline
\hspace*{3mm}channel\hspace*{3mm} 
  & \hspace*{3mm}state\hspace*{3mm}
  & \hspace*{3mm}$E_{\Lambda\Lambda}$ (MeV) \hspace*{3mm}
  & \hspace*{3mm}$P_{\Lambda\Lambda}$ (\%)\hspace*{3mm} 
  & \hspace*{3mm}$P_{\Xi N}$ (\%)\hspace*{3mm}
  & \hspace*{3mm}$P_{\Sigma\Sigma}$ (\%)\hspace*{3mm}  \\
\hline
$\Lambda\Lambda$ 
  & $0^+_1$ & 1.18 & 100.0 & $-$ & $-$ \\
\hline
$\Lambda\Lambda$-$\Xi N$ 
  & $0^+_1$ & 1.10 & 99.9 & 0.1 & $-$ \\
$\Lambda\Lambda$-$\Sigma\Sigma$ 
  & $0^+_1$ & $-21.4$ & 48.0 & $-$ & 52.0 \\
  & $0^+_2$ &  1.40  & 99.7 & $-$ & 0.3 \\
$\Xi N$-$\Sigma\Sigma$ 
  & $0^+_1$ & 4.41   & $-$ & 31.2 & 68.8 \\
  & $0^+_2$ & 29.34  & $-$ & 99.9 & 0.1 \\
\hline
$\Lambda\Lambda$-$\Xi N$-$\Sigma\Sigma$
  & $0^+_1$ & $-1475$ & 26.6 & 31.1 & 42.4 \\
  & $0^+_2$ & 1.15 & 100.0 & 0.0 & 0.0 \\
\hline
\hline
\multicolumn{6}{c}{NSC97f potential}\\
\hline
channel & state
  & $E_{\Lambda\Lambda}$ (MeV) 
  & $P_{\Lambda\Lambda}$ (\%) & $P_{\Xi N}$ (\%) & $P_{\Sigma\Sigma}$ (\%)  \\
\hline
$\Lambda\Lambda$ 
  & $0^+_1$ & 1.19 & 100.0 & $-$ & $-$ \\
\hline
$\Lambda\Lambda$-$\Xi N$ 
  & $0^+_1$ & 1.15 & 100.0 & 0.0 & $-$ \\
$\Lambda\Lambda$-$\Sigma\Sigma$ 
  & $0^+_1$ & 1.08 & 99.9 & $-$ & 0.1 \\
$\Xi N$-$\Sigma\Sigma$ 
  & $0^+_1$ & $-81.98$   & $-$ & 28.2 & 71.8 \\
  & $0^+_2$ & 29.31  & $-$ & 100.0 & 0.0 \\
\hline
$\Lambda\Lambda$-$\Xi N$-$\Sigma\Sigma$
  & $0^+_1$ & $-1624$ & 23.9 & 32.1 & 44.0 \\
  & $0^+_2$ & 1.17 & 100.0 & 0.0 & 0.0 \\
\hline
\hline
\end{tabular}
\end{center}
\end{table}

\clearpage
\begin{table}
\caption{
Calculated $\Delta B_{\Lambda\Lambda}$ 
and the channel components of {$^{~6}_{\Lambda\Lambda}$He} 
with (without) the exclusion operator $V_{PBS}$ in Eq.~(\ref{PBS}),
where the NSC97e and NSC97f potentials are used.
}
\label{tab:2}
\begin{center}
\begin{tabular}{cccccccc}
\hline
\hline
\hspace*{5mm}potential\hspace*{5mm} & \hspace*{5mm}channel\hspace*{5mm} 
        & \hspace*{2mm}$V_{PBS}$\hspace*{2mm} & state 
        & \hspace*{5mm}$\Delta B_{\Lambda\Lambda}$ (MeV)\hspace*{5mm} 
        & \hspace*{3mm}$P_{\Lambda\Lambda}$ (\%)\hspace*{3mm} 
        & \hspace*{3mm}$P_{\Xi N}$ (\%)\hspace*{3mm}
        & \hspace*{3mm}$P_{\Sigma\Sigma}$ (\%)\hspace*{3mm} \\
\hline
          & $\Lambda\Lambda$-$\Xi N$-$\Sigma\Sigma$ & {\it Yes}  & $0^+_1$ & 0.61 & 99.77 & 0.21 & 0.01 \\ 
          & $\Lambda\Lambda$                                & {\it No} & $0^+_1$ & 0.33 & 100 & $-$ & $-$ \\
NSC97e & $\Lambda\Lambda$-$\Xi N$                      & {\it No}  & $0^+_1$ & 1.36 & 99.45 & 0.55 & $-$ \\
          & $\Lambda\Lambda$-$\Xi N$-$\Sigma\Sigma$ & {\it No}  & $0^+_1$ & 1467 & 26.65 & 30.97 & 42.38 \\ 
          & $\Lambda\Lambda$-$\Xi N$-$\Sigma\Sigma$ & {\it No}  & $0^+_2$ & 0.61 & 99.77 & 0.21 & 0.01 \\ 
\hline
          & $\Lambda\Lambda$-$\Xi N$-$\Sigma\Sigma$ & {\it Yes} & $0^+_1$ & 0.36 & 99.81 & 0.18 & 0.01 \\ 
          & $\Lambda\Lambda$                               & {\it No} & $0^+_1$ & 0.09 & 100 & $-$ & $-$ \\
NSC97f & $\Lambda\Lambda$-$\Xi N$                      & {\it No} & $0^+_1$ & 0.56 & 99.79 & 0.21 & $-$ \\
          & $\Lambda\Lambda$-$\Xi N$-$\Sigma\Sigma$ & {\it No} & $0^+_1$ & 1614 & 24.01 & 31.98 & 44.01 \\ 
          & $\Lambda\Lambda$-$\Xi N$-$\Sigma\Sigma$ & {\it No} & $0^+_2$ & 0.36 & 99.81 & 0.18 & 0.01 \\ 
\hline
\hline
\end{tabular}
\end{center}
\end{table}

\clearpage
\begin{figure}
\begin{center}
\includegraphics*[scale=0.5]{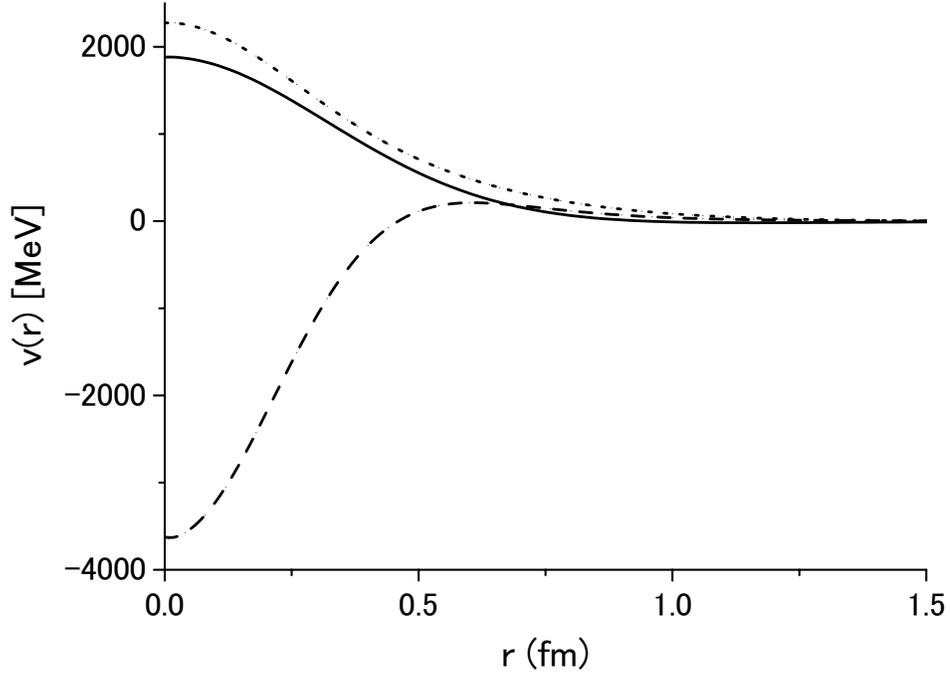}
\caption{
Radial behavior of the bare potentials, 
 $\upsilon_{\Lambda\Lambda-\Lambda\Lambda}(r)$ (solid line),
 $\upsilon_{{\Xi N}-{\Xi N}}(r)$ (dotted) and
 $\upsilon_{\Lambda\Lambda-{\Xi N}}(r)$ (dashed), 
 for the NSC97e potential. 
}
\label{fig:1}
\end{center}
\end{figure}

\clearpage
\begin{figure}
\begin{center}
\includegraphics*[scale=0.5]{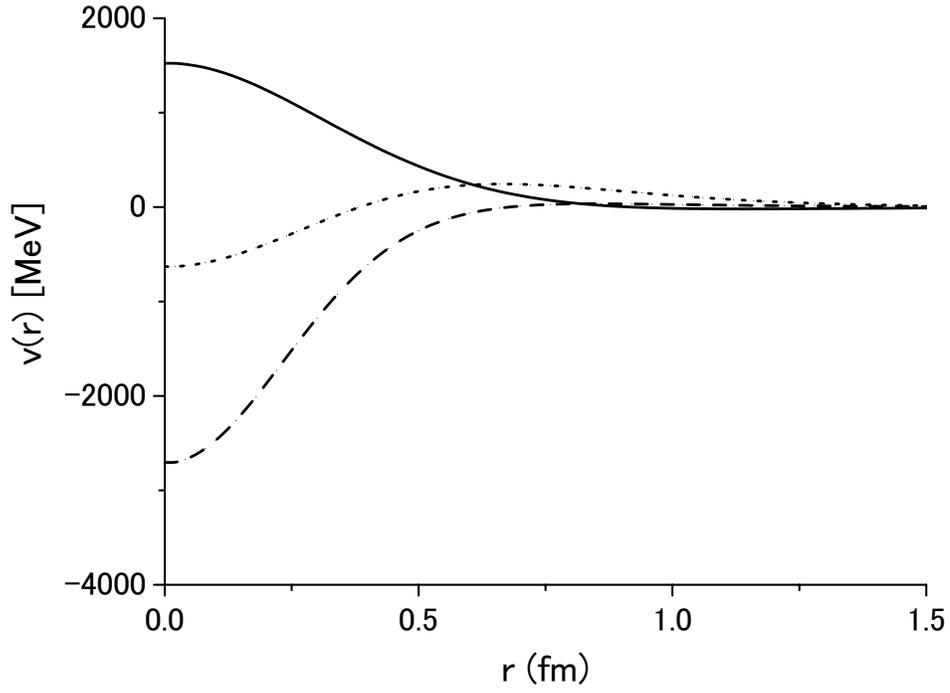}
\caption{
Radial behavior of the bare potentials, 
 $\upsilon_{\Lambda\Lambda-\Lambda\Lambda}(r)$ (solid line),
 $\upsilon_{{\Sigma\Sigma}-{\Sigma\Sigma}}(r)$ (dotted) and
 $\upsilon_{\Lambda\Lambda-{\Sigma\Sigma}}(r)$ (dashed), 
 for the NSC97e potential. 
}
\label{fig:2}
\end{center}
\end{figure}

\clearpage
\begin{figure}
\begin{center}
\includegraphics*[scale=0.5]{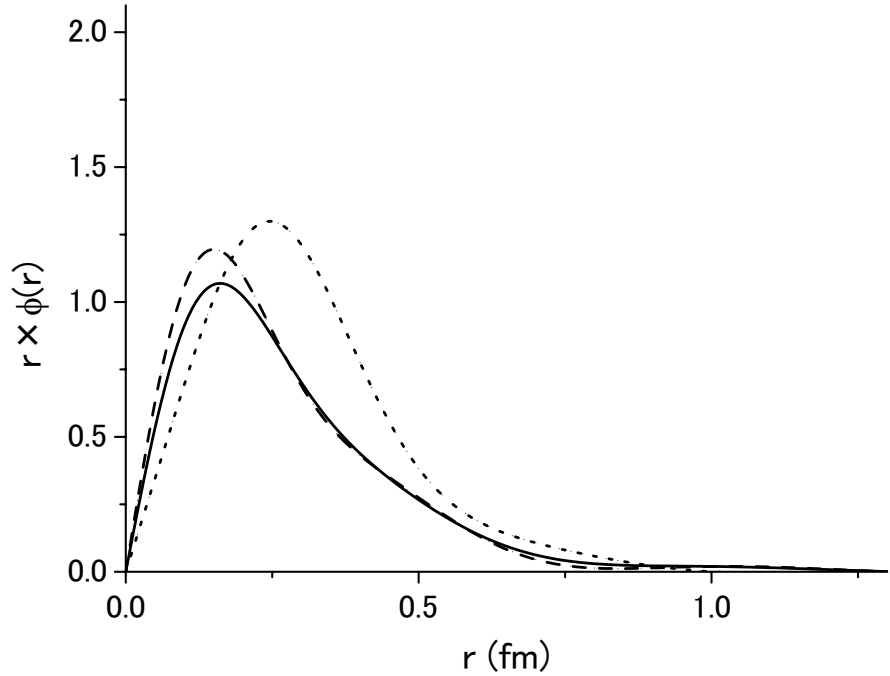}
\caption{
Radial behavior of the relative wave function (multiplied to $r$)
 between the two baryons in the pseudo bound state with 
 the flavor-SU(3)-$\{8s\}$-like character, where we
 use the NSC97e potential with the momentum-dependent term.
The solid, dashed and dotted lines denote the $\Lambda\Lambda$-,
 $\Xi N$- and $\Sigma\Sigma$-channel wave functions, respectively.
}
\label{fig:3}
\end{center}
\end{figure}

\clearpage
\begin{figure}
\begin{center}
\includegraphics*[scale=0.5]{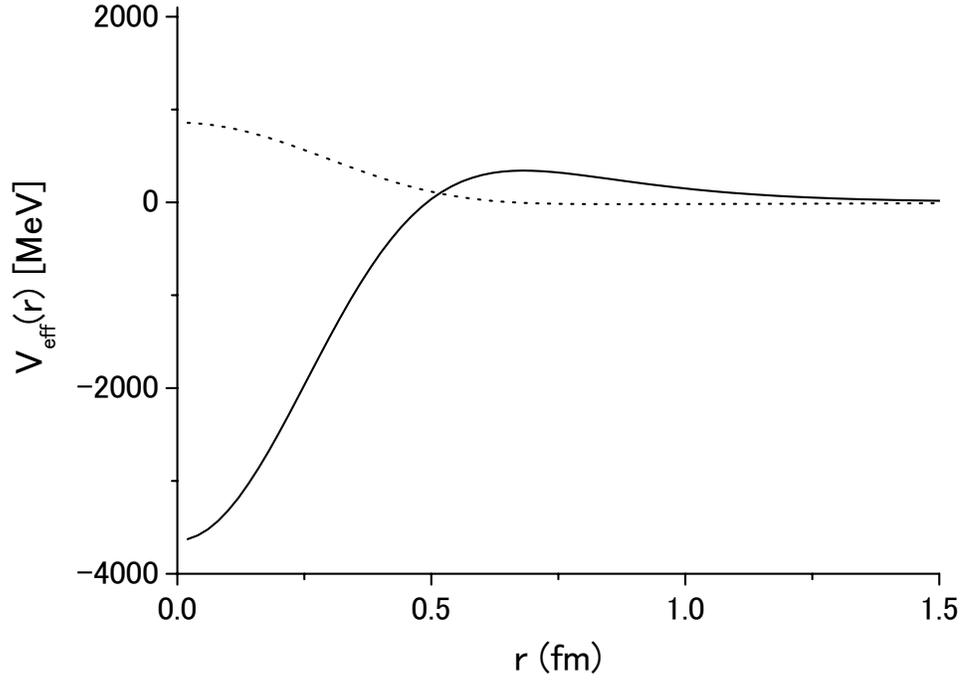}
\caption{
Effective potentials for the flavor-SU(3) states ($^1S_0$)
 with $S=-2$ and $I=0$:~$\{8s\}$ (solid line) and $\{27\}$ (dotted),
 where we use the NSC97e potential with the momentum-dependent term.
}
\label{fig:4}
\end{center}
\end{figure}

\end{document}